# Lorentz Skew Scattering Nonreciprocal Magneto-Transport

Xiu Fang Lu[1,9], Xue-Jin Zhang[2,9], Naizhou Wang[1], Jin Cao[3], Dan Zhao[4], Hui Wang[1], Tao Wu[4], Xian Hui Chen[4], Shen Lai[2], Cong Xiao[5*], Shengyuan A. Yang[3*], and Weibo Gao[1,6,7,8*]

[1]Division of Physics and Applied Physics, School of Physical and Mathematical Sciences, Nanyang Technological University, Singapore 637371, Singapore.

[2]Institute of Applied Physics and Materials Engineering, Faculty of Science and Technology, University of Macau, Macau SAR, China.

[3]Research Laboratory for Quantum Materials, Department of Applied Physics, The Hong Kong Polytechnic University, Hong Kong, China

[4]Department of Physics and Hefei National Laboratory for Physical Science at Microscale, University of Science and Technology of China, Hefei, Anhui 230026, China.

[5]Interdisciplinary Center for Theoretical Physics and Information Sciences (ICTPIS), Fudan University, Shanghai 200433, China

[6]School of Electrical and Electronic Engineering, Nanyang Technological University, Singapore, Singapore

[7]Centre for Quantum Technologies, National University of Singapore, Singapore, Singapore

[8]National Centre for Advanced Integrated Photonics (NCAIP) Singapore, Nanyang Technological University, Singapore, Singapore

[9]These authors contributed equally: Xiu Fang Lu, Xue-Jin Zhang

E-mail: wbgao@ntu.edu.sg, shengyuan.yang@polyu.edu.hk, congxiao@fudan.edu.cn

**Abstract**

**In materials with broken inversion symmetry, nonreciprocal magneto-transport (NRMT) manifests as a bilinear dependence of charge conductivity on applied electric (*E*) and magnetic (*B*) fields. This phenomenon is deeply rooted in symmetry and electronic quantum geometry, holding promise for novel rectification and detector technologies. Existing experimental studies generally attribute NRMT to Zeeman-driven mechanisms and exhibit quadratic scaling with conductivity. Here, we report a previously unknown NRMT microscopic mechanism - Lorentz skew scattering (LSK) - revealed through the discovery of an unprecedented quartic scaling law of NRMT as well as quantitative agreement between theory and experiment in BiTeBr. LSK emerges from the interplay of Lorentz force and skew scattering, bridging classical field effect to quantum scattering effect on the Fermi surface. We demonstrate that the LSK dominates NRMT in BiTeBr, and elucidate that this dominance over other possible contributions stems from high mobility and strong Rashba splitting. The finding of LSK mechanism is of unique importance because it unveils the leading NRMT effect in high-mobility systems and suggests a universal principle towards strong NRMT by enhancing electronic relaxation time in topological materials, rendering a new designing idea for low-dissipation rectifiers and high-performance quantum electronics.**

**Introduction**

In terms of current response, NRMT corresponds to the nonreciprocal component $j^{NLMT} = \chi E^2 B$ [1-18]. It is nonreciprocal, since it leads to a difference in response when the field (or current) direction is reversed. Such a phenomenon was first studied in materials with chiral structures and was known as electrical magneto-chiral anisotropy[1,19]. Lately, it was actively explored also in achiral systems lacking inversion symmetry[14,20-25], such as surfaces and interfaces of bulk crystals with central asymmetric lattices; and it came under the names of bilinear magnetoelectric resistance (for longitudinal response, $j \parallel E$) or nonlinear planar Hall effect (for transverse response, $j \perp E$). In these studies, the magnetic field is typically applied within the transport plane, such that characteristic features of the response tensor $\chi$ may be extracted from angular dependence of the in-plane magnetic field. It was found that the NRMT often manifests intriguing quantum geometrical properties of Bloch electrons[2,18,26]. In Weyl semimetals, NRMT may arise from chiral anomaly[27] or Berry curvature mechanisms[28]. NRMT results detected in other systems were usually attributed to chiral spin textures due to spin-orbit coupling, and the effect of magnetic field is to deform the Fermi surface via Zeeman coupling to magnetic moment of electrons[3-8,29-31]. The experimentally reported NRMT so far is generally weak, and existing understanding of microscopic mechanisms for NRMT is far from complete, both severally hindering its potential applications.

A common approach to distinguish different mechanisms of nonlinear transport effects is to perform scaling analysis, which examines how the response tensor $\chi$ depends on the longitudinal conductivity $\sigma_{xx}$ [32-48]. Experimentally, this is usually achieved by varying the temperature of measurement, and the resulting scaling relation usually takes a polynomial form: $\chi = \sum_{i=0,1,2,\ldots} c_i \sigma_{xx}^i$. As for NRMT in nonmagnetic systems, the Hall response may have an intrinsic contribution with $i = 0$ [16,49]; whereas for longitudinal response, the previously proposed mechanisms, including the Zeeman-coupling mechanism, is invariably dominated by the $\chi \propto \sigma_{xx}^2$ scaling [3-8,29-31]. Interestingly, a recent theory proposed a new mechanism for NRMT - the Lorentz skew scattering (LSK)[50], the schematics of which is shown in Fig. 1a. Distinct from other mechanisms, in LSK, the magnetic field enters via Lorentz force, rather than Zeeman coupling[4-8,15,31] or other quantum mechanical effects[21,27,28], which directly affects real-space motion of electrons instead of modifying momentum-space band structures. In the meantime, the quantum geometry of Bloch electrons is encoded in the skew scattering process, whose scattering rate is shown to be related to the Berry curvature on Fermi surface[50]. Most importantly, LSK gives a quartic scaling behaviour $\propto \sigma_{xx}^4$, sharply distinct from all previous NRMT mechanisms. It is shown to be the leading degree term in the scaling relation, which should dominate the NRMT response in systems with high mobility. Since low-dissipation electronic devices commonly desire high-mobility active materials, the LSK mechanism offers a new strategy to enhance NRMT (and rectification) effect for applications. Nevertheless, till now, such distinctive quartic scaling behaviour associated to LSK has not been reported in experiment yet.

From the above discussion, a key to observe LSK-dominated NRMT is to have samples with high mobility (the meaning of "high" is explained later). And for Lorentz force to take effect in the planar setup, one should use a 3D system rather than surfaces or interfaces. We achieve this goal in *n*-doped polar semiconductor BiTeBr with high mobility. The scaling analysis reveals for the first time the predicted quartic scaling behaviour corresponding to LSK. Our experimental result further finds quantitative agreement with theoretical calculations, confirming LSK as the dominant mechanism of the observed nonlinear response.

**Setup and nonlinear planar transport measurement**

As shown in Fig. 1b, BiTeBr exhibits threefold rotational symmetry within *ab*-plane and a polar axis along the *c*-axis. The ABC stacking of Te, Bi and Br triangular atomic layers breaks both the inversion symmetry and the horizontal mirror symmetry, which fulfils the requirement of NRMT with in-plane magnetic field. To determine the crystal axes of our BiTeBr flake, we employed the optical second harmonic generation (SHG) technique. As depicted in Fig. 1c, the SHG intensity displays a distinct threefold rotational symmetry, with the direction of minimum (maximum) intensity corresponding to the crystal axis marked as $x$ ($y$). Here, $y$ is the direction within the vertical mirror planes. After determining the crystal axis, we fabricated a high quality, few layer BiTeBr device with the thickness of about 15 nm and 12 radically distributed electrodes (Device #1) (see methods for device details). Our initial step involved assessing the contact quality of these electrodes (See supplementary Fig. S1 for the two-terminal *I-V* curves for all 12 electrodes), which demonstrated good contact across all electrodes. We then carried out the basic characterization on Device #1 (Supplementary Fig. S2). The electron carrier density obtained from Hall measurement is ~ $1.0 \times 10^{19}$ cm$^{-3}$, which is similar to previous studies[4]. However, the extracted linear resistivity $\rho_{xx}$ (varying from 0.30 mΩcm at 2 K to ~ 1.28 mΩcm at 300 K) is almost an order of magnitude smaller than previously reported values, indicating the high quality of our samples[4,46].

To measure the linear and nonlinear electrical transport, an AC current ($I^\omega$) at a fixed frequency $\omega$=17.777 Hz was applied in the *ab*-plane of the device, which makes an angle $\varphi$ with respect to the crystal $x$ axis (Fig. 1d). We first characterized the first harmonic voltage drops at longitudinal ($V_\parallel$) and transverse ($V_\perp$) directions under zero magnetic field, as shown in Fig. 1e ($\varphi = 0$). Both $V_\parallel$ and $V_\perp$ increase linearly with current, and the magnitude of $V_\perp$ remains negligible compared to $V_\parallel$, indicating good contact and proper alignment of electrodes. Next, we measured the NRMT response of Device #1. The measurement configuration is illustrated in Fig. 1d. We applied a magnetic field within the *ab*-plane of BiTeBr, whose direction described by the polar angle $\theta$ may vary in plane (details of the electrical measurement shown in Methods). We then measured the second-harmonic response in both longitudinal ($V_\parallel^{2\omega}$) and transverse ($V_\perp^{2\omega}$) directions. Fig. 1f shows the results of Device #1 as a function of the relative angle ($\theta - \varphi$), with a 5 T in-plane magnetic field and a 200 μA driving current at 50 K. To obtain the response signal induced by magnetic field, we subtracted the field-independent background of $V_\parallel^{2\omega}$ and $V_\perp^{2\omega}$ (Supplementary Fig. S3). As depicted in Fig. 1f, $V_\parallel^{2\omega}$ and $V_\perp^{2\omega}$ exhibits almost perfect sine and cosine dependence with a 90° phase shift. Additionally, both $V_\parallel^{2\omega}$ and $V_\perp^{2\omega}$ flip signs when reversing the direction of the magnetic field. To characterize the magnitude of response, we use the notation $\Delta V_\perp^{2\omega}$ ($\Delta V_\parallel^{2\omega}$) to represent the amplitude of the oscillating curve of $V_\perp^{2\omega}$ ($V_\parallel^{2\omega}$). When we changed the orientation of driving current direction, the waveforms of $V_\perp^{2\omega}$ and $V_\parallel^{2\omega}$ remain invariant as functions of ($\theta - \varphi$) (Supplementary Fig. S4), and the amplitudes $\Delta V_\perp^{2\omega}$ and $\Delta V_\parallel^{2\omega}$ remains nearly constant, as presented in Fig. 1g.

At a fixed current direction, we measured the $V_\parallel^{2\omega}$ and $V_\perp^{2\omega}$ at 50 K under different magnitudes of driving current and magnetic field, as shown in Fig. 2 (the data of $V_\parallel^{2\omega}$ are shown in Supplementary Fig. S5). One can see that $\Delta V_\perp^{2\omega}$ increases with $I^\omega$ (Fig. 2a), and scales with the square of the linear longitudinal voltage $V_\parallel$ (Fig. 2b), i.e., $\Delta V_\perp^{2\omega} \propto (V_\parallel)^2$. Furthermore, for a fixed $I^\omega$, $\Delta V_\perp^{2\omega}$ increases linearly with the magnetic field up to at least 9 T (Fig. 2c-d), showing $\Delta V_\perp^{2\omega} \propto B$. Similar results are also observed in the longitudinal response (Supplementary Fig. S5). These results unambiguously demonstrate that the observed nonlinear signals $\Delta V_\perp^{2\omega}$ and $\Delta V_\parallel^{2\omega}$ correspond to the NRMT response, which is $\propto E^2 B$.

This identification of measured signal with NRMT also perfectly explains the observed angular dependence of $V_\parallel^{2\omega}$ and $V_\perp^{2\omega}$. Constrained by the $C_{3v}$ point group symmetry of BiTeBr, the longitudinal and transverse NRMT responses have angular dependence in the outer product form $\sim \boldsymbol{I} \times \boldsymbol{H} \cdot \hat{\boldsymbol{P}}$ and inner product form $\sim \boldsymbol{I} \cdot \boldsymbol{H}$, respectively. Here, $\hat{\boldsymbol{P}}$ is the unit vector along the polar axis ($z$-axis). This accounts for the observed sine and cosine dependence: $V_\parallel^{2\omega} \sim \Delta V_\parallel^{2\omega} \sin(\theta - \varphi)$ and $V_\perp^{2\omega} \sim \Delta V_\perp^{2\omega} \cos(\theta - \varphi)$. The $C_{3v}$ symmetry ensures that the amplitudes $\Delta V_\perp^{2\omega}$ and $\Delta V_\parallel^{2\omega}$ remains nearly constant regardless of the current injection direction. Such a clean angular dependence on the relative angle of the driving current and magnetic field is unique to in-plane field, which also helps exclude the possibility of out-of-plane field component due to misalignment.

**Observation of quartic scaling**

Having identified the NRMT response, we next investigated the underlying physical mechanism by the scaling analysis. To perform this, we measured $V_\perp^{2\omega}$ and $V_\parallel^{2\omega}$ across varying magnitudes of electric and magnetic fields at different temperatures. Fig. 3a shows the extracted nonlinear transverse response field $\Delta E_\perp^{2\omega}$ as a function of longitudinal electric field $E_\parallel$ at different temperatures under a 5 T in-plane magnetic field. Here, $E_\parallel \equiv V_\parallel/L_\parallel$ and $\Delta E_\perp^{2\omega} \equiv \Delta V_\perp^{2\omega}/L_\perp$, with $L_\perp$ and $L_\parallel$ being the transverse and longitudinal lengths of the device, respectively. As depicted in Fig. 3a, $\Delta E_\perp^{2\omega}$ scales linearly with $(E_\parallel)^2$ at all temperatures, and the slopes of $\Delta E_\perp^{2\omega}$ - $(E_\parallel)^2$ curves decrease monotonically with increasing temperature. In addition to the 5 T data, we also measured $\Delta E_\perp^{2\omega}$ - $(E_\parallel)^2$ under varying magnetic fields from 1 T to 9 T, which exhibit the same features. We plot all the measured values of $\Delta E_\perp^{2\omega}/(E_\parallel)^2$ in Fig. 3b, which shows a consistent decrease with temperature across all magnetic fields. The longitudinal response $\Delta E_\parallel^{2\omega}$ exhibits similar behaviours, as shown in Supplementary Fig. S6.

For scaling analysis, we focus on the NRMT response coefficient $\chi_\perp^{2\omega} = \sigma \frac{\Delta E_\perp^{2\omega}}{(E_\parallel)^2 B}$ and $\chi_\parallel^{2\omega} = \sigma \frac{\Delta E_\parallel^{2\omega}}{(E_\parallel)^2 B}$. Relating them to the rank-4 NRMT response tensor defined by $j_i^{2\omega} = \chi_{ijk\ell}^{2\omega} E_j^\omega E_k^\omega B_\ell$, where all indices $\in \{x, y\}$ and repeated indices are summed over, one finds that $\chi_\perp^{2\omega} = |\chi_{yxxx}^{2\omega}|$ and $\chi_\parallel^{2\omega} = |\chi_{xxxy}^{2\omega}|$ in our setup. As discussed, previously reported mechanisms for NRMT give a $\chi \propto \sigma_{xx}^2$ scaling[3-6,8,31]. The $\sigma_{xx}$ as a function of temperature is presented in Fig. 3c (due to $C_{3v}$ symmetry, the linear conductivity is isotropic in the plane). We plot $\chi_\perp^{2\omega}$ (Fig. 3d) and $\chi_\parallel^{2\omega}$ (Fig. S6d) versus $\sigma_{xx}^2$. One can see clearly that the NRMT does not conform to the quadratic scaling relation for conventional mechanisms. It is evident that both the transverse and longitudinal responses are dominated by contributions of higher powers in $\sigma_{xx}$, which has not been observed before. This indicates some new mechanism must be in action here.

By carefully examining the scaling behaviour of the data, we find the curves can be perfectly fit by a simple quartic scaling relation $\chi = c\sigma_{xx}^4$, with vanishingly small intercepts, as shown in Fig. 3e and 3f. This clean result signifies the overwhelming dominance of the quartic scaling contribution in both the transverse and longitudinal NRMT measured here. Moreover, the magnitudes of the two responses are quite similar, with the slope $\chi_\perp^{2\omega}/\sigma_{xx}^4$ being $0.33 \times 10^{-16}$ AV$^{-2}$T$^{-1}$S$^{-4}$cm$^4$ and $\chi_\parallel^{2\omega}/\sigma_{xx}^4$ being $0.38 \times 10^{-16}$ AV$^{-2}$T$^{-1}$S$^{-4}$cm$^4$. Since the crystal symmetry does not require these values to coincide, the common quartic scaling behaviour and the close magnitudes suggest that the observed transverse and longitudinal nonreciprocal transport should share the same physical origin.

To explore the physical origin of the unconventional quartic scaling behaviour, we first ascertained that the temperature variation of longitudinal conductivity comes from the variation in electronic relaxation time $\tau$, i.e., the mobility instead of the carrier density. We measured the Hall resistivity (under perpendicular magnetic fields) and extracted the carrier density of our BiTeBr device across a temperature range from 2 to 300 K (Supplementary Fig. S2). The perfect linearity of Hall resistivity indicates a single band (single-$\tau$) dominated transport in our BiTeBr device (which is also confirmed by our first-principles result). The carrier density of our BiTeBr device keeps almost unchanged across all temperatures, proving that the temperature variation in $\sigma_{xx}$ comes from the change in $\tau$. (Indeed, as shown in Fig. 3c, the mobility and $\sigma_{xx}$ exhibit nearly the same temperature dependence.) Therefore, the quartic scaling behaviour in Fig. 3e and 3f further shows that $\chi_\perp^{2\omega} \propto \tau^4$ and $\chi_\parallel^{2\omega} \propto \tau^4$ regarding the temperature dependence.

We have also repeated the measurement and scaling analysis for another BiTeBr Device #2, which has a different thickness (~20 nm) and carrier density (~$2.5 \times 10^{19}\ cm^{-3}$ at 2 K). The data from Device #2, shown in Supplementary Fig. S7 to S9, exhibit consistent behaviours as those of Device #1, demonstrating that the NRMT and quartic scaling observed here are indeed physical and reproducible.

**Lorentz skew scattering mechanism**

The observed quartic scaling $\chi \propto \tau^4$ is consistent with the LSK mechanism. As shown in Ref. 50, the LSK gives a NRMT current expressed as $\boldsymbol{j}^{\text{LSK}} = -e\sum_l g_l^{\text{LSK}} \boldsymbol{v}_l$, where $\boldsymbol{v}_l$ is the velocity of a Bloch electron in state $l$, and

$$g_l^{\text{LSK}} = -\tau^4 \left[\widehat{D}_E\{\widehat{D}_B, \hat{I}_{sk}\} + \widehat{D}_B\{\hat{I}_{sk}, \widehat{D}_E\} + \hat{I}_{sk}\{\widehat{D}_E, \widehat{D}_B\}\right]\widehat{D}_E f^0 \qquad (1)$$

is the LSK off-equilibrium distribution function. Here, $f^0$ is the Fermi distribution, $\hat{I}_{sk}$ is the skew-scattering collision-integral operator, whose action on a distribution function $f_l$ is given by $\hat{I}_{sk} f_l = -\sum_{l'} \omega_{l'l}^{3a}(f_l + f_{l'})$, with $\omega_{l'l}^{3a}$ being the skew scattering rate (details in Methods). $\widehat{D}_E = -e\boldsymbol{E} \cdot \partial_{\boldsymbol{k}}$ and $\widehat{D}_B = -e\boldsymbol{v}_l \times \boldsymbol{B} \cdot \partial_{\boldsymbol{k}}$ are differential operators corresponding to driving terms of electric force and

Lorentz force in the Boltzmann kinetic equation, and $\{\hat{D}_E, \hat{D}_B\} \equiv \hat{D}_E\hat{D}_B + \hat{D}_B\hat{D}_E$. In the presence of both impurity and phonon scattering, the relaxation time $\tau$ is contributed by both, and $g_l^{LSK} \propto \tau^4$ leads to the quartic scaling behaviour of the nonreciprocal transport. It was shown that the LSK mechanism gives the highest degree term in the scaling relation[50], so it is expected to dominate the response for high-mobility samples, which is the case here. As shown in Fig. 3c, the mobility rises from 500 $cm^2V^{-1}s^{-1}$ at 300 K to about 2000 $cm^2V^{-1}s^{-1}$ at 20 K. We note that in linear anomalous transport, it has been well established that the skew scattering contribution dominates over other contributions in 3d transition metals Fe, Co, and Ni when $\sigma_{xx} \geq 10^6 \, \Omega^{-1}cm^{-1}$, which is termed as "high conductivity regime"[51]. Since the carrier densities of these metals are about $10^{23} \, cm^{-1}$, the mobilities of these high conductivity systems are smaller than $100 \, cm^2V^{-1}s^{-1}$. Therefore, the mobility in our BiTeBr device (other devices also possess mobilities of similar magnitudes) can indeed be viewed as high in NRMT, benefitting the observation of LSK transport.

To perform quantitative evaluation based on the above theoretical formula, we first calculate the band structures of BiTeBr by first-principles calculation (Supplementary Fig. S10). The result shows that the low-energy physics is well described by the following 3D Rashba model around $A$ point of the Brillouin zone:

$$\mathcal{H} = \frac{1}{2m_{xy}}\left(k_x^2 + k_y^2\right) + \frac{k_z^2}{2m_z} + \alpha_R(k_y\sigma_x - k_x\sigma_y) + \frac{1}{2}(k_+^3 + k_-^3)(\lambda\sigma_z + \lambda_0 k_z) \qquad (2)$$

where the first two terms are usual quadratic dispersing terms, the third term is a Rashba-type spin-orbit coupling, and the last term is a warping term which ensures the model recovers the $C_{3v}$ symmetry of the system. The model parameters are obtained by fitting the first-principles band structure (see Methods). This 3D Rashba model features a Weyl point at $k = 0$, i.e., the $A$ point,. The Fermi level position estimated from the carrier density value is quite close to the Weyl point, as shown in Fig. 4a. Notably, in this region, due to the strong Rashba spin-orbit coupling strength $\alpha_R$, the density of states for the inner Rashba band is strongly suppressed. Our estimation shows that the density of states of outer Rashba band is an order of magnitude larger than the inner one. This indicates that the transport is likely dominated by the outer band, which explains the observed single-band transport behaviour. Moreover, a previous study on BiTeBr[46] with similar carrier densities has found that Coulomb impurities with a screening wavevector much smaller than the Fermi-wavevector difference between the two Rashba bands is the dominant scattering source. This further supports the independent-band transport picture and the dominance of the outer Rashba band in NRMT.

The identification of transport being dominated by the outer Rashba band also helps exclude other NRMT mechanisms of quartic scaling, suggesting that the LSK is left as the uniquely predominant candidate. A recent theory on the second order electrical nonlinear transport $\propto E^2$ in magnetic systems predicted quartic scaling from the compositions of two skew scattering processes (SKSK)[35]. These terms are also possible here if the nonmagnetic material is viewed as being "magnetised" by the applied magnetic field. We now inspect this possibility in our devices. The ratio between LSK and the Zeeman corrected SKSK contribution is approximately $\frac{\chi^{\text{LSK}}}{\chi^{\text{SKSK}}} \sim \sigma\omega_c\tau / \frac{c_1\Delta}{\hbar v_F k_F}\sigma\frac{g\mu_B B}{\Delta} \sim \frac{2\hbar v_F k_F \tau}{\hbar g c_1}$, where $\omega_c$ is the cyclotron frequency, $\mu_B$ is the Bohr magneton, $g$ is the $g$-factor, and $\Delta$ is the vertical interband separation. $\hbar v_F k_F$ is a characteristic energy scale related to electronic motion on the Fermi surface, which takes the place of Fermi energy in systems with strong spin-orbit coupling[46]. $c_1$ is a dimensionless factor measuring the ratio of the time scales of lowest Born scattering and skew scattering, which is much less than unity[39, 52, 53]. The estimated relaxation times according to the measured mobility at 40 K are large, reaching $\tau \simeq 0.15$ ps for Device#1 and $\tau \simeq 0.1$ ps for Device#2 (Supplementary Fig. S11). In addition, the ratio $\frac{\chi^{\text{LSK}}}{\chi^{\text{SKSK}}}$ is further enhanced by the large Fermi velocity and Fermi wavevector of the outer Rashba band ($\hbar v_F k_F \simeq 290$ meV in Device#1 and $\hbar v_F k_F \simeq 450$ meV in Device#2). Therefore, assume a considerable value of $c_1 = 0.1$ and the $g$-factor $g = 60^4$, the LSK is more than one order of magnitude greater than the SKSK in our devices, giving the overwhelmingly dominant quartic-scaling contribution.

Moreover, adopting the above effective model and modelling skew scattering processes from the screened Coulomb impurities, we were able to estimate the LSK induced NRMT coefficient $\chi$ (details in Methods). The results are shown in Fig. 4b and 4c along with the experimental result. The agreement is excellent. The theoretical result reproduces not only the temperature dependence of $\chi$ but also its correct order of magnitude. For example, at 100 K, the theoretical estimation for Device#1 (Device#2) gives $\chi_\perp^{2\omega} \sim 3.04 \times 10^{-4}$ AV$^{-2}$T$^{-1}$ ($\sim 1.41 \times 10^{-4}$ AV$^{-2}$T$^{-1}$), which is close to the experimental value of $\chi_\perp^{2\omega} \sim 3.77 \times 10^{-4}$ AV$^{-2}$T$^{-1}$ ($\sim 1.55 \times 10^{-4}$ AV$^{-2}$T$^{-1}$). Such quantitative agreement offers strong support that LSK is the dominant mechanism underlying the observed NRMT response here.

**Conclusion**

We have reported a distinct quartic scaling behaviour of NRMT response in high-mobility BiTeBr samples. Via systematic analysis, we reveal that such unusual nonreciprocal transport originates from a novel LSK mechanism, in which the magnetic field enters via Lorentz force rather than Zeeman coupling and the quantum geometric character manifests in the skew scattering rate. The topological band structures on the n-doping side help boost the mobility of electrons which makes LSK gain dominance due to its high scaling power in relaxation time. The strong Rashba spin-orbit coupling in BiTeBr helps make a single-band transport scenario. However, it should be noted that the LSK mechanism itself does not require spin-orbit coupling, so it may also be observed in materials with negligible spin-orbit coupling strength. Due to its quartic scaling behaviour, the LSK induced nonreciprocal transport should be greatly enhanced in materials with long electronic relaxation time and strong Berry curvatures on Fermi surface. This implies that topological materials, such as graphene superlattices and Weyl semimetals (where $\tau$ may reach 10 to 100 ps), could be good platforms to achieve large NRMT. Our work thus suggests a new route to giant transport nonreciprocity in high-mobility materials with low dissipation and power consumption, which holds great promise for efficient nonlinear devices such as rectifiers and diodes.

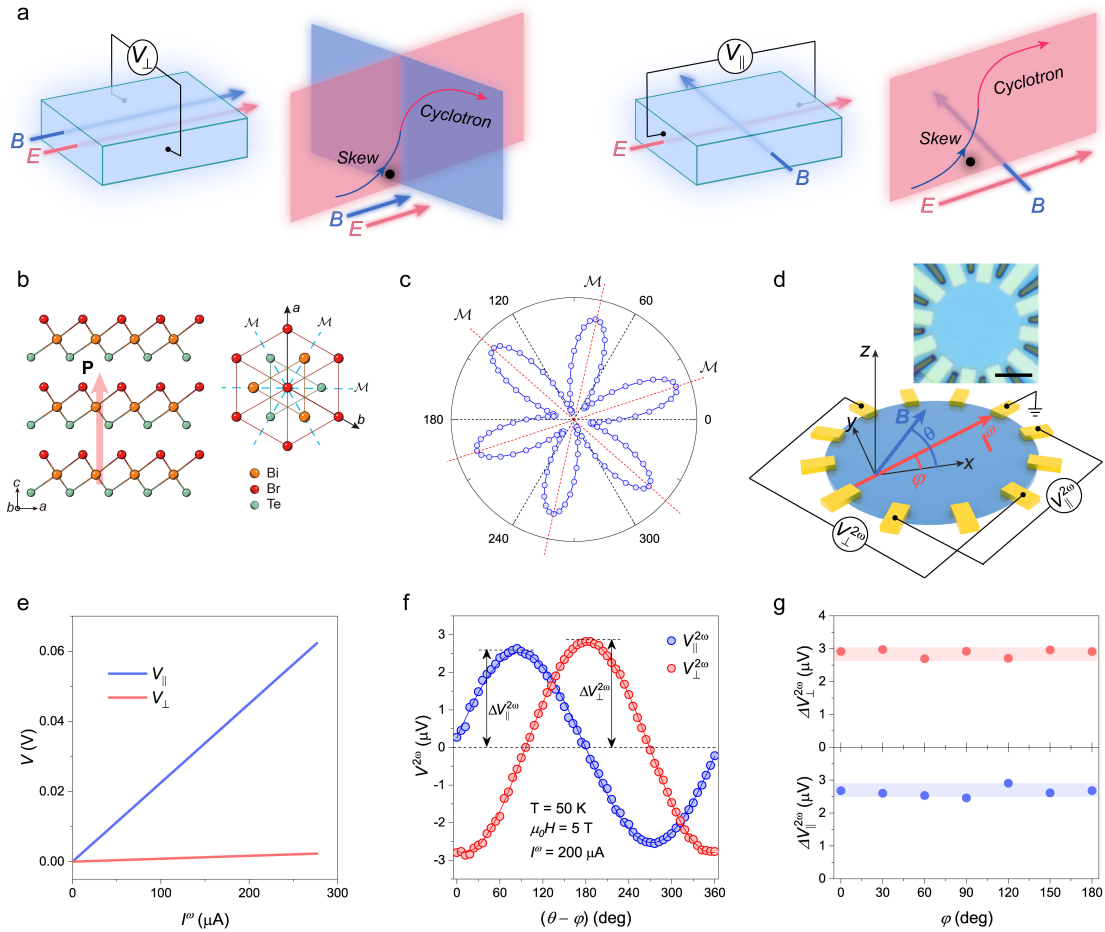

**Fig. 1 | Schematics of LSK, symmetry properties of BiTeBr structure, and the measurement of NLMT**. **a**, Schematics of the microscopic processes of LSK. Left panel: Transverse LSK transport in the configuration of parallel fields. Right panel: Longitudinal LSK transport in the configuration of perpendicular fields. **b**, Side and top views of the BiTeBr crystal structure. There is a spontaneous electrical polarization along *c*-axis due to the inversion symmetry breaking and threefold rotational symmetry in the *ab*-plane, as shown. **c**, The angle-dependent optical second harmonic generation (SHG) intensity of an exfoliated BiTeBr flake. The direction of minimum intensity is identified as the crystal axis. **d**, Schematic illustration of the measurement configuration for NLMT. The *x, y, z*-axes represent the crystal axis direction, the direction transverse to the crystal axis, and the out-of-plane directions, respectively. An AC current ($I^\omega$) is applied to device with an angle $\varphi$ to the *x*-axis, while an in-plane magnetic field *B* makes an angle $\theta$ with respect to the *x*-axis. The second harmonic response is measured simultaneously in the longitudinal ($V_\parallel^{2\omega}$) and transverse ($V_\perp^{2\omega}$) direction. Insert: optical image of Device #1. scale bar: 5 μm. **e**, The linear longitudinal and transverse voltage, $V_\parallel$ and $V_\perp$, as a function of $I^\omega$ at 50K under zero magnetic field. **f**, The nonlinear responses $V_\parallel^{2\omega}$ and $V_\perp^{2\omega}$ as a function of angle ($\theta - \varphi$), measured with the configuration illustrated in (**d**). **g**, Amplitude of the nonlinear responses, $\Delta V_\parallel^{2\omega}$ and $\Delta V_\perp^{2\omega}$, as a function of driving current orientation $\varphi$.

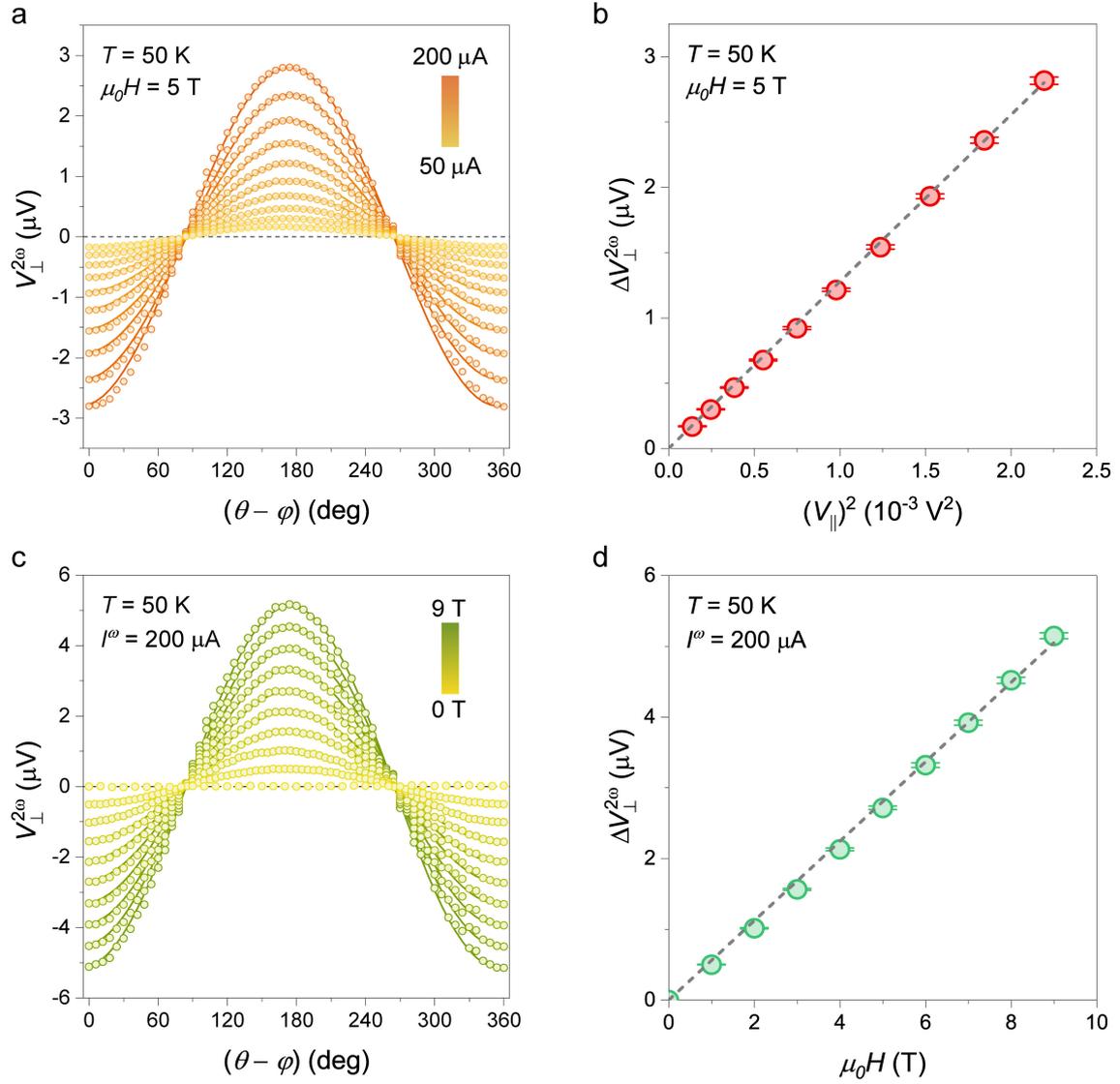

**Fig. 2 | The NLMT under varying electric and magnetic fields. a**, and **b**, The nonlinear planar transverse response $V_\perp^{2\omega}$ as a function of angle $(\theta - \varphi)$, measured with varying $I^\omega$ from 50 to 200 μA (**a**) and the corresponding fitted amplitude of $V_\perp^{2\omega} \sim (\theta - \varphi)$ cosine curve, $\Delta V_\perp^{2\omega}$ (with standard error bars) as a function of the square of linear longitudinal voltage $V_\parallel$ (**b**). **c**, and **d**, $V_\perp^{2\omega}$ as a function of angle $(\theta - \varphi)$, measured under varying in-plane magnetic field $(\mu_0 H)$ ranging from 1 to 9T (**c**) and the corresponding fitted $\Delta V_\perp^{2\omega}$ (with standard error bars) scaling with $\mu_0 H$ (**d**).

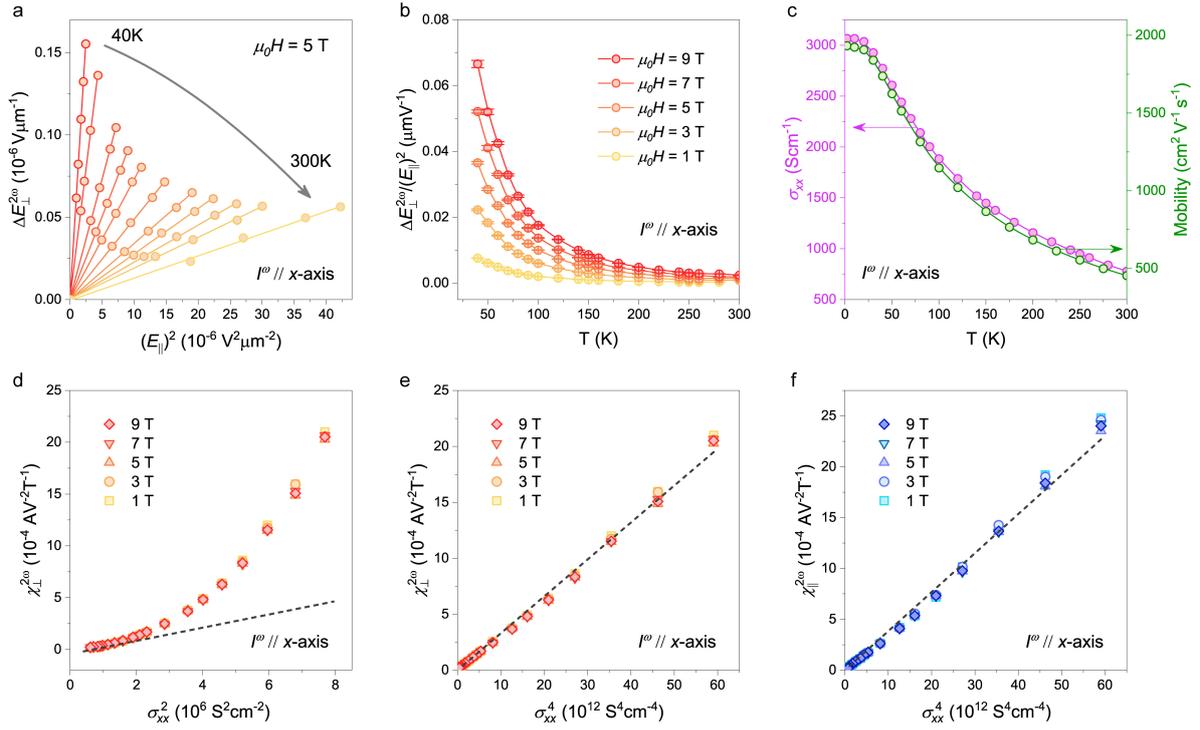

**Fig. 3 | Temperature dependence and scaling law of the NLMT observed in BiTeBr**. **a**, $\Delta E_\perp^{2\omega}$ as a function of $(E_\parallel)^2$ at temperatures ranging from 40K to 300K, measured under a 5 T in-plane magnetic field with $I^\omega$ applied along $x$-axis. The slope of $\Delta E_\perp^{2\omega}$ - $(E_\parallel)^2$ decreases as temperature increases. **b**, $\Delta E_\perp^{2\omega}/(E_\parallel)^2$ as a function of temperature, measured under varying magnetic field from 1 to 9 T. **c**, Temperature dependence of longitudinal conductivity $\sigma_{xx}$ and mobility. **d**, Scaling the NLMT coefficient $\chi_\perp^{2\omega}$ as $\sigma_{xx}^2$. **e** and **f**, Scaling the NLMT coefficient $\chi_\perp^{2\omega}$ (**e**) and $\chi_\parallel^{2\omega}$ (**f**) as $\sigma_{xx}^4$.

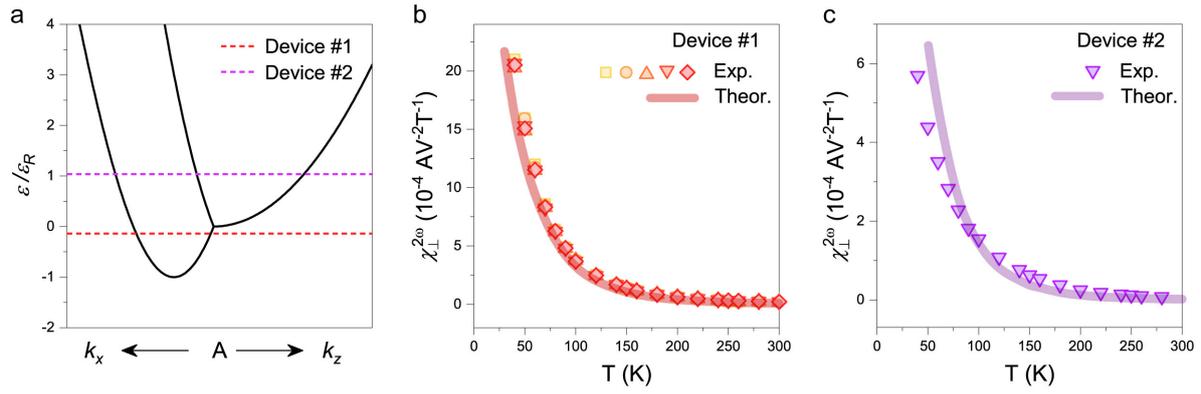

**Fig. 4 | The LSK as the microscopic physical origin of the observed NLMT in BiTeBr. a,** Band structure of the Rashba model fitted from the first-principles bands of BiTeBr. The red (purple) dashed line represents the Fermi level of Device#1 (Device#2). Here $\varepsilon_R = \frac{\alpha_R^2}{4t} = 42$ meV is the Rashba energy. **b** and **c** show the NLMT coefficient $\chi_\perp^{2\omega}$ of Device#1 and Device#2, respectively. The scatters represent experimental data, whereas the curves depict theoretical results of LSK effect.

## Methods

### Device fabrication

The BiTeBr flakes were mechanically exfoliated on a SiO$_2$/Si substrate inside nitrogen-filled glovebox with O$_2$ and H$_2$O level kept below 1 ppm. The crystalline axes of the exfoliated BiTeBr flakes were determined using the second harmonic generation (SHG) method. To prevent degradation, a PMMA polymer layer was coated on BiTeBr flakes inside glove box before SHG measurement. To enhance electrode contact, the stencil mask method was employed to deposit metal contacts (Cr/Au, 2/50nm) on the samples. To standardize the shape of the device and avoid geometrical complications in the current path, two methods were used: reactive ion etching with SF$_6$ and CHF$_3$ gases, and direct scraping with the tip of an atomic force microscope. After all fabrication processes, a PMMA polymer layer was applied once again to prevent degradation before measurement.

### Nonlinear planar transport measurements

The nonlinear planar transport measurements were conducted in a Quantum Design PPMS cryostat using a Horizontal Rotator, allowing the BiTeBr devices to be rotated through a full 360° in the presence of an applied magnetic field. A harmonic current was applied to the BiTeBr devices during the measurement, and second harmonic voltage drops in both longitudinal and transverse directions were recorded using a standard lock-in technique (Zurich MFLI). An in-plane magnetic field ranging from 1 to 9T was applied throughout the measurements, with the device being rotated by the Horizontal Rotator around its *x*-axis at an angle θ relative to the magnetic field direction. The phases of the first- and second- harmonic signals was confirmed to be approximately 0° and 90°, respectively, during all electrical measurements.

### First-principles calculations

The first-principles calculations are based on the density functional theory (DFT), using the projector-augmented wave method[54] as implemented in the Vienna ab-initio simulation package (VASP)[55,56]. The Perdew-Burke-Ernzerhof (PBE) functional was utilized to capture exchange-correlation effects[57]. The cutoff energy is 400 eV. The lattice structures are relaxed with an energy convergence criterion of $10^{-8}$ eV and a force convergence criterion of 0.01 eV/Å. The Brillouin zone is sampled by a Γ-centered Monkhorst-Pack k-point mesh[58] with size 15 × 15 × 7 for self-consistent calculations. Spin-orbit coupling is included in BiTeBr calculations. Wannier tight-binding model with p orbitals of Bi, Te and Br atoms was constructed by using the Wannier90 package[59-61].

## Skew scattering in three-dimensional Rashba bands

The BiTeBr has $C_{3v}$ point group symmetry generated by a three-fold rotation $C_{3z}$ and mirror symmetry $\mathcal{M}_x$. Its low-energy Hamiltonian can be captured by a three-dimensional Rashba model

$$\mathcal{H} = t(k_x^2 + k_y^2) + t_z k_z^2 + \alpha_R(k_y \sigma_x - k_x \sigma_y) + (k_+^3 + k_-^3)(\frac{\lambda \sigma_z}{2} + \frac{\lambda_0 k_z}{2}),$$

measured from A point. Here, $k_\pm = k_x \pm i k_y$, and $\sigma$ denotes Pauli matrix. The third term represents Rashba spin-orbit coupling, $\lambda$ term added hexagonal warping of the Fermi surface, and $\lambda_0$ term deforms the hexagonal warping to a trigonal warping at nonzero $k_z$ as required by $C_{3z}$. In our calculations, we take model parameters as $t = 25$ eVÅ$^2$, $t_z = 5$ eVÅ$^2$, $\alpha_R = 2.0$ eVÅ, $\lambda = 50$ eVÅ$^3$, $\lambda_0 = 20$ eVÅ$^4$, which can reproduce the band structures from first-principles calculations.

The energy dispersion of this model is

$$\epsilon(k) \approx t_z k_z^2 + t(k_x^2 + k_y^2) + \eta \sqrt{\lambda^2 (k_x^3 - k_x k_y^2)^2 + \alpha_R(k_x^2 + k_y^2)} + \lambda_0 (k_x^3 - k_x k_y^2) k_z,$$

where $\eta = \pm 1$, and the corresponding eigenstates are

$$|u_k^+\rangle = \begin{pmatrix} \cos(\theta/2) \\ -i \sin(\theta/2) e^{i\phi} \end{pmatrix}, |u_k^-\rangle = \begin{pmatrix} -i \sin(\theta/2) \\ \cos(\theta/2) e^{i\phi} \end{pmatrix}.$$

Here, $k_\perp = \sqrt{k_x^2 + k_y^2}$, $\tan \phi = \frac{k_y}{k_x}$, and $\cos \theta = \frac{\lambda k_\perp^2 \cos 3\phi}{\sqrt{\alpha_R^2 + (\lambda k_\perp^2 \cos 3\phi)^2}}$. The leading term of the Bloch state overlap is

$$\langle u_k^+ | u_{k'}^+ \rangle = \langle u_{k'}^- | u_k^- \rangle \approx \frac{1}{2}(1 + e^{i(\phi' - \phi)}) + \frac{\lambda}{4\alpha_R}(k_\perp^2 \cos 3\phi + k_\perp'^2 \cos 3\phi')(1 - e^{i(\phi' - \phi)}) + \mathcal{O}(\lambda^2),$$

Consider the screened Coulomb scattering (Gaussian-natural units) $V_{kk'} = \frac{4\pi e^2 \alpha}{(k-k')^2 + q_{TF}^2} \langle u_k | u_{k'} \rangle$, where $\alpha^{-1} \approx 16$ is the relative dielectric constant of BiTeBr, we get the lowest Born scattering rate

$$\omega_{k,k'}^{\pm(2)} = \frac{2\pi}{\hbar} \langle V_{kk'} V_{k'k} \rangle_{dis} \delta(\epsilon_k - \epsilon_{k'})$$

$$= \frac{2\pi}{\hbar} n_i \left( \frac{4\pi e^2 \alpha}{(k-k')^2 + q_{TF}^2} \right)^2 (1 + \cos(\phi' - \phi)) \delta(\epsilon_k - \epsilon_{k'}) + \mathcal{O}(\lambda),$$

and the antisymmetric scattering rate responsible for skew scattering

$$\omega_{k,k'}^{\pm(3a)} = \frac{4\pi^2}{\hbar} \sum_{k''} \Im \langle V_{kk''} V_{k''k'} V_{k'k} \rangle_{dis} \delta(\epsilon_k - \epsilon_{k''}) \delta(\epsilon_k - \epsilon_{k'})$$

$$= \mp \frac{\lambda}{2\pi \hbar \alpha_R} \delta(\epsilon_k - \epsilon_{k'}) \int d\mathbf{k''} [\sin(\phi' - \phi) k_\perp''^2 \cos 3\phi'' + \sin(\phi'' - \phi') k_\perp^2 \cos 3\phi$$

$$+\sin(\phi - \phi'')k'^2_\perp \cos 3\phi']\frac{4\pi e^2 \alpha}{(\boldsymbol{k}-\boldsymbol{k}')^2 + q^2_{TF}}\frac{4\pi e^2 \alpha}{(\boldsymbol{k}'-\boldsymbol{k}'')^2 + q^2_{TF}}\frac{4\pi e^2 \alpha}{(\boldsymbol{k}''-\boldsymbol{k})^2 + q^2_{TF}}\delta(\epsilon_{\boldsymbol{k}} - \epsilon_{\boldsymbol{k}''})$$

for the outer (−) and inner (+) Rashba bands. Substituting the above equations into the LSK distribution function [Eq. (1) of the main text] we can get the expressions for the LSK nonlinear current. Given the complexity of high-dimensional numerical integrals, we adopt semiquantitative estimation to get meaningful expressions for the LSK nonlinear conductivity. The characteristic Coulomb potential is approximated as $\frac{4\pi e^2 \alpha}{k_F^2}$, then the LSK nonlinear conductivity at the second harmonic can be simplified into

$$\chi^{2\omega,\text{LSK}} \approx \frac{2^{11}\pi^5 \tau^4 e^{10} \alpha^3 n_i t^2 \lambda \lambda_0}{\hbar^6 \alpha_R}\frac{D_F^3}{k_F}.$$

Here, $D_F$ is the density of states at the Fermi level. The factor $\frac{D_F^3}{k_F}$ for the outer Rashba band is 77 (170) times of that for the inner band in Device#1 (Device#2), which is consistent with the conclusion drawn from qualitative model analysis and out-of-plane Hall measurement, supporting also the validity of our estimation procedure. Assuming that the charged impurity density is approximately the same as the carrier density, we then get the theoretical results shown in Fig. 4b-c.

**Data availability:** All data shown in this paper are available from the corresponding author upon reasonable request. Source data are provided with this paper.

**Acknowledgements**: This work is supported by ASTAR (M21K2c0116, M24M8b0004), Singapore National Research foundation (NRF-CRP22-2019-0004, NRF-CRP30-2023-0003, NRF2023-ITC004-001 and NRF-MSG-2023-0002) and Singapore Ministry of Education Tier 2 Grant (MOE-T2EP50221-0005, MOE-T2EP50222-0018). C.X. was sponsored by National Natural Science Foundation of China (grant no.12574114) and the start-up funding from Fudan University. S.A.Y. was supported by The HK PolyU Start-up Grant No. (P0057929).

**Author contributions**: W.G., S.A.Y., and C.X. conceived and supervised the project. X.F.L. fabricated the devices and performed the transport and SHG measurements with help from N.W. C.X., X.-J.Z., J.C., and H.W. performed theoretical calculation. X.F.L. and C.X. analysed the data. D.Z., T. W., and X.H.C. grew the BiTeBr single crystals. X.F.L., N.W., C.X, S.A.Y, and W.G. wrote the paper with input from all authors.

**Competing interests:** Authors claim there are no competing interests.

**References**

1   Rikken, G. L., Folling, J. & Wyder, P. Electrical magnetochiral anisotropy. *Phys. Rev. Lett.* **87**, 236602 (2001).

2   Tokura, Y. & Nagaosa, N. Nonreciprocal responses from non-centrosymmetric quantum materials. *Nat. Commun.* **9**, 3740 (2018).

3   Ideue, T. & Iwasa, Y. Symmetry breaking and nonlinear electric transport in van der Waals nanostructures. *Annual Review of Condensed Matter Physics* **12**, 201-223 (2021).

4   Ideue, T. *et al.* Bulk rectification effect in a polar semiconductor. *Nat. Phys.* **13**, 578-583 (2017).

5   Zhang, S. S.-L. & Vignale, G. in *Spintronics XI*.  97-107 (SPIE).

6   He, P. *et al.* Bilinear magnetoelectric resistance as a probe of three-dimensional spin texture in topological surface states. *Nat. Phys.* **14**, 495-499 (2018).

7   He, P. *et al.* Nonlinear Planar Hall Effect. *Phys. Rev. Lett.* **123**, 016801 (2019).

8   He, P. *et al.* Observation of out-of-plane spin texture in a $SrTiO_3$(111) two-dimensional electron gas. *Phys. Rev. Lett.* **120**, 266802 (2018).

9   Rikken, G. & Avarvari, N. Strong electrical magnetochiral anisotropy in tellurium. *Phys. Rev. B* **99**, 245153 (2019).

10  Guillet, T. *et al.* Observation of Large Unidirectional Rashba Magnetoresistance in Ge(111). *Phys. Rev. Lett.* **124**, 027201 (2020).

11  Dyrdal, A., Barnas, J. & Fert, A. Spin-momentum-locking inhomogeneities as a source of bilinear magnetoresistance in topological insulators. *Phys. Rev. Lett.* **124**, 046802 (2020).

12  Li, Y. *et al.* Nonreciprocal charge transport up to room temperature in bulk Rashba semiconductor alpha-GeTe. *Nat. Commun.* **12**, 540 (2021).

13  Calavalle, F. *et al.* Gate-tuneable and chirality-dependent charge-to-spin conversion in tellurium nanowires. *Nat. Mater.* **21**, 526-532 (2022).

14  Legg, H. F. *et al.* Giant magnetochiral anisotropy from quantum-confined surface states of topological insulator nanowires. *Nat. Nanotechnol.* **17**, 696-700 (2022).

15  Zhang, Y. *et al.* Large magnetoelectric resistance in the topological Dirac semimetal alpha-Sn. *Sci. Adv.* **8**, eabo0052 (2022).

16  Huang, Y. X., Feng, X., Wang, H., Xiao, C. & Yang, S. A. Intrinsic Nonlinear Planar Hall Effect. *Phys. Rev. Lett.* **130**, 126303 (2023).

17  Hu, M.-W. et al. Modulation of chiral anomaly and bilinear magnetoconductivity in Weyl semimetals by impurity resonance states. *Phys. Rev. B* **109**, 155154 (2024).

18  Suárez-Rodríguez, M. et al. Symmetry origin and microscopic mechanism of electrical magnetochiral anisotropy in tellurium. *Phys. Rev. B* **111**, 024405 (2025).

19  Pop, F., Auban-Senzier, P., Canadell, E., Rikken, G. L. & Avarvari, N. Electrical magnetochiral anisotropy in a bulk chiral molecular conductor. *Nat. Commun.* **5**, 3757 (2014).

20  Wang, Y. *et al.* Large bilinear magnetoresistance from Rashba spin-splitting on the surface of a topological insulator. *Phys. Rev. B* **106**, L241401 (2022).

21  Wang, Y. *et al.* Nonlinear transport due to magnetic-field-induced flat bands in the nodal-line semimetal $ZrTe_5$. *Phys. Rev. Lett.* **131**, 146602 (2023).

22  Wang, N. *et al.* Non-centrosymmetric topological phase probed by non-linear Hall effect. *Natl. Sci. Rev.* **11**, nwad103 (2024).

23  Zhang, X. *et al.* Light-induced giant enhancement of nonreciprocal transport at $KTaO_3$-based interfaces. *Nat. Commun.* **15**, 2992 (2024).

24  Li, C. *et al.* Observation of giant non-reciprocal charge transport from quantum Hall states in a topological insulator. *Nat. Mater.* **23**, 1208-1213 (2024).

25  Min, L. *et al.* Colossal room-temperature non-reciprocal Hall effect. *Nat. Mater.* **23**, 1671-1677 (2024).

26  Zhao, T.-Y. et al. Magnetic Field Induced Quantum Metric Dipole in Dirac Semimetal $Cd_3As_2$. *Phys. Rev. Lett.* **135**, 026601 (2025).

27  Morimoto, T. & Nagaosa, N. Chiral anomaly and giant magnetochiral anisotropy in noncentrosymmetric Weyl semimetals. *Phys. Rev. Lett.* **117**, 146603 (2016).

28  Yokouchi, T., Ikeda, Y., Morimoto, T. & Shiomi, Y. Giant magnetochiral anisotropy in Weyl semimetal $WTe_2$ induced by diverging Berry curvature. *Phys. Rev. Lett.* **130**, 136301 (2023).

29  Okumura, S., Tanaka, R. & Hirobe, D. Chiral orbital texture in nonlinear electrical conduction. *Phys. Rev. B* **110**, L020407 (2024).

30  He, P. *et al.* Nonlinear magnetotransport shaped by Fermi surface topology and convexity. *Nat. Commun.* **10**, 1290 (2019).

31  Tuvia, G. *et al.* Enhanced nonlinear response by manipulating the Dirac point at the (111) $LaTiO_3/SrTiO_3$ interface. *Phys. Rev. Lett.* **132**, 146301 (2024).

32  Hou, D. *et al.* Multivariable scaling for the anomalous Hall effect. *Phys. Rev. Lett.* **114**, 217203 (2015).

33  Du, Z., Wang, C., Li, S., Lu, H.-Z. & Xie, X. Disorder-induced nonlinear Hall effect with time-reversal symmetry. *Nat. Commun.* **10**, 3047 (2019).

34  Du, Z., Lu, H.-Z. & Xie, X. Nonlinear hall effects. *Nat. Rev. Phys.* **3**, 744-752 (2021).

35  Huang, Y.-X., Xiao, C., Yang, S. A. & Li, X. Scaling law and extrinsic mechanisms for time-reversal-odd second-order nonlinear transport. *Phys. Rev. B* **111**, 155127 (2025).

36  Kang, K., Li, T., Sohn, E., Shan, J. & Mak, K. F. Nonlinear anomalous Hall effect in few-layer $WTe_2$. *Nat. Mater.* **18**, 324-328 (2019).

37  Lai, S. *et al.* Third-order nonlinear Hall effect induced by the Berry-connection polarizability tensor. *Nat. Nanotechnol.* **16**, 869-873 (2021).

38  Kumar, D. *et al.* Room-temperature nonlinear Hall effect and wireless radiofrequency rectification in Weyl semimetal $TaIrTe_4$. *Nat. Nanotechnol.* **16**, 421-425 (2021).

39  He, P. *et al.* Quantum frequency doubling in the topological insulator $Bi_2Se_3$. *Nat. Commun.* **12**, 698 (2021).

40  Tiwari, A. *et al.* Giant c-axis nonlinear anomalous Hall effect in $Td$-$MoTe_2$ and $WTe_2$. *Nat. Commun.* **12**, 2049 (2021).

41  Ma, T. *et al.* Growth of bilayer $MoTe_2$ single crystals with strong non-linear Hall effect. *Nat. Commun.* **13**, 5465 (2022).

42  Duan, J. *et al.* Giant second-order nonlinear Hall effect in twisted bilayer graphene. *Phys. Rev. Lett.* **129**, 186801 (2022).

43  Sinha, S. *et al.* Berry curvature dipole senses topological transition in a moiré superlattice. *Nat. Phys.* **18**, 765-770 (2022).

44  Wang, N. *et al.* Quantum-metric-induced nonlinear transport in a topological antiferromagnet. *Nature* **621**, 487-492 (2023).

45  Gao, A. *et al.* Quantum metric nonlinear Hall effect in a topological antiferromagnetic heterostructure. *Science* **381**, 181-186 (2023).

46  Lu, X. F. *et al.* Nonlinear transport and radio frequency rectification in BiTeBr at room temperature. *Nat. Commun.* **15**, 245 (2024).

47  Liao, X. *et al.* Nonlinear valley and spin valves in bilayer graphene. *Phys. Rev. Appl.* **22**, 054078 (2024).

48  Wang, A.-Q. *et al.* Orbital anomalous Hall effect in the few-layer Weyl semimetal $TaIrTe_4$. *Phys. Rev. B* **110**, 155434 (2024).

49  Huang, Y.-X. *et al.* Nonlinear current response of two-dimensional systems under in-plane magnetic field. *Phys. Rev. B* **108**, 075155 (2023).

50  Xiao, C., Huang, Y.-X. & Yang, S. A. Lorentz Skew Scattering and Giant Nonreciprocal Magneto-Transport. *arXiv preprint arXiv:2411.07746* (2024).

51  Nagaosa, N., Sinova, J., Onoda, S., MacDonald, A. H. & Ong, N. P. Anomalous Hall effect. *Rev. Mod. Phys.* **82**, 1539-1592 (2010)

52  Isobe, H., Xu, S.-Y. & Fu, L. High-frequency rectification via chiral Bloch electrons. *Sci. Adv.* **6**, eaay2497 (2020).

53  He, P. et al. Graphene moiré superlattices with giant quantum nonlinearity of chiral Bloch electrons. *Nat. Nanotech.* **17**, 378-383 (2022).

54  Blöchl, P. E. Projector augmented-wave method. *Phys. Rev. B* **50**, 17953 (1994).

55  Kresse, G. & Hafner, J. Ab initio molecular-dynamics simulation of the liquid-metal-amorphous-semiconductor transition in germanium. *Phys. Rev. B* **49**, 14251 (1994).

56  Kresse, G. & Furthmüller, J. Efficient iterative schemes for ab initio total-energy calculations using a plane-wave basis set. *Phys. Rev. B* **54**, 11169 (1996).

57  Perdew, J. P., Burke, K. & Ernzerhof, M. Generalized gradient approximation made simple. *Phys. Rev. Lett.* **77**, 3865 (1996).

58  Monkhorst, H. J. & Pack, J. D. Special points for Brillouin-zone integrations. *Phys. Rev. B* **13**, 5188 (1976).

59  Marzari, N. & Vanderbilt, D. Maximally localized generalized Wannier functions for composite energy bands. *Phys. Rev. B* **56**, 12847 (1997).

60  Souza, I., Marzari, N. & Vanderbilt, D. Maximally localized Wannier functions for entangled energy bands. *Phys. Rev. B* **65**, 035109 (2001).

61  Mostofi, A. A. *et al.* wannier90: A tool for obtaining maximally-localised Wannier functions. *Comput. Phys. Commun.* **178**, 685-699 (2008).